%% file: sbes.tex
\documentclass[sigconf]{acmart}
\input{packages}

\newcolumntype{P}[1]{>{\centering\arraybackslash}p{#1}}
\newcolumntype{M}[1]{>{\centering\arraybackslash}m{#1}}

\newcommand{\zup}{\textsc{Zup}\xspace}

\newcommand{\totalPessoasCompletaramTreinamentos}{16\xspace}
\newcommand{\totalPessoasNaoCompletaramTreinamentos}{24\xspace}

\newcommand{\MyBox}[1]{\vspace{3mm}\noindent\framebox[\columnwidth][c]{\parbox[b]{0.95\columnwidth}{ #1 }}\vspace{3mm}}

\definecolor{newcolor}{rgb}{1.0,0.49,0.0}

\AtBeginDocument{%
  \providecommand\BibTeX{{%
    \normalfont B\kern-0.5em{\scshape i\kern-0.25em b}\kern-0.8em\TeX}}}

\setcopyright{acmcopyright}
\copyrightyear{2023}
\acmYear{2023}
\acmDOI{XXXXXXX.XXXXXXX}

\acmConference[SBES 2023]{The 37th Brazilian Symposium on Software Engineering -- Education Track}{25 - 27 September, 2023}{Campo Grande, Brasil}

\acmPrice{15.00}
\acmISBN{978-1-4503-XXXX-X/18/06}

\begin{document}

\title{Large Language Models for Education: \\ Grading Open-Ended Questions Using ChatGPT}

\author{Gustavo Pinto}
\affiliation{%
  \institution{UFPA \& Zup Innovation}
  \city{Belém}
  \state{PA}
  \country{Brazil}
}
\email{gustavo.pinto@zup.com.br}

\author{Isadora Cardoso-Pereira}
\affiliation{%
  \institution{Zup Innovation}
  \city{Recife}
  \country{Brazil}
}
\email{isadora.cardoso@zup.com.br}

\author{Danilo Monteiro Ribeiro}
\affiliation{%
  \institution{Zup Innovation}
  \city{Recife}
  \country{Brazil}
}
\email{danilo.ribeiro@zup.com.br}

\author{Danilo Lucena}
\affiliation{%
  \institution{Zup Innovation}
  \city{João Pessoa}
  \country{Brazil}
}
\email{danilo.ucena@zup.com.br}

\author{Alberto de Souza}
\affiliation{%
  \institution{Zup Innovation}
  \city{São Paulo}
  \country{Brazil}
}
\email{alberto.tavares@zup.com.br}

\author{Kiev Gama}
\affiliation{%
  \institution{Universidade Federal de Pernambuco}
  \city{Recife}
  \country{Brazil}
}
\email{kiev@cin.ufpe.br}


\renewcommand{\shorttitle}{Grading of Open-Ended Questions Using ChatGPT}
\renewcommand{\shortauthors}{Pinto et al.}

\begin{abstract}

As a way of addressing increasingly sophisticated problems, software professionals face the constant challenge of seeking improvement. However, for these individuals to enhance their skills, their process of studying and training must involve feedback that is both immediate and accurate. In the context of software companies, where the scale of professionals undergoing training is large, but the number of qualified professionals available for providing corrections is small, delivering effective feedback becomes even more challenging. To circumvent this challenge, this work presents an exploration of using Large Language Models (LLMs) to support the correction process of open-ended questions in technical training.

In this study, we utilized ChatGPT to correct open-ended questions answered by 42 industry professionals on two topics. Evaluating the corrections and feedback provided by ChatGPT, we observed that it is capable of identifying semantic details in responses that other metrics cannot observe. Furthermore, we noticed that, in general, subject matter experts tended to agree with the corrections and feedback given by ChatGPT.
\end{abstract}


\begin{CCSXML}
<ccs2012>
   <concept>       
       <concept_id>10011007.10011074.10011075.10011078</concept_id>
       <concept_desc>Software and its engineering~Software design tradeoffs</concept_desc>
       <concept_significance>500</concept_significance>
       </concept>
 </ccs2012>
\end{CCSXML}

\ccsdesc[500]{Software and its engineering~Software design tradeoffs}

\keywords{ChatGPT, Open-ended Questions, Automated grading}


\maketitle

\section{Introduction}

The software industry regularly presents challenges to development teams. Companies and teams continually strive to improve operational efficiency while reducing costs and maintaining or increasing productivity. In this context, software developers must continuously enhance their skills to remain relevant in their careers~\cite{noviceBegel}. The process of studying and training\footnote{For the context of this work, we use the terms ``training'' and ``studying'' interchangeably.} is integral to developers' work routine, supporting their continuous improvement and playing a crucial role in enhancing technical skills.

During the training process, problem-solving activities or exercises are essential for two main reasons. Firstly, they help solidify and self-assess the understanding of new concepts. Secondly, they indicate whether the individual's current knowledge is sufficient for performing tasks~\cite{tomporowski2015exercise, smith1982learning, makransky2016virtual}.
The feedback that developers receive regarding their exercises is just as important as solving them~\cite{vollmeyer2005surprising, thurlings2013understanding}. Good feedback is crucial for the learning process of software developers, as it provides insights into their progress and areas for improvement. However, offering effective feedback can be challenging, requiring broad subject knowledge, availability, and the ability to identify knowledge gaps~\cite{gielen2010improving}. Moreover, providing feedback promptly adds to the complexity of this task~\cite{thurlings2013understanding}.

In the context of a software producing organization, it is necessary for the individuals who provide feedback to find time to do that. Thus, they need to allocate hours they usually dedicate for tasks related to software development and allocate part of their schedule for this grading and feedback activity, which reduce their operational efficiency. This can have a significant impact, especially for companies that invest in their employees' learning. This is specifically the scenario experienced at \textsc{Zup Innovation}.

Due to \zup's size, with several thousand employees, of which approximately 90\% work in engineering teams, 
theoretical exams predominantly consist of closed-ended questions, either single-choice or multiple-choice. 
However, the usage of closed-ended questions presents a significant limitation in the evaluation process, given their limited \emph{feedback}.
 %Ademais, a avaliação de questões abertas é frequentemente demandada, embora seja um desafio para o time de Educação. 
Aiming to accelerate the grading and feedback process, this work presents an investigation into the use of ChatGPT as a supplementary evaluation method. Our goal is to understand whether ChatGPT could be considered as a mechanism for grading open-ended questions in the training process employed by \zup.

To accomplish this, the study began by creating a set of open-ended questions on two topics of interest to \zup: (1) web application caching and (2) stress and performance testing. We asked two experts in this field to answer three questions each. With the responses from these experts, we conducted a pilot experiment involving six developers from the engineering team at \zup. In this pilot, we administered an online questionnaire and asked the developers to respond to the six open-ended questions. We used ChatGPT to correct the pilot questions and evaluated the quality of the prompts for answer grading. After receiving feedback from the pilot participants, we randomly invited 100 more people from the engineering team, of whom 50 had completed at least one technical training, while the other group had started but not completed any training. Of these, \totalPessoasCompletaramTreinamentos and \totalPessoasNaoCompletaramTreinamentos individuals from each group, respectively, completed the questionnaire (N=40). In this work, we bring the following contributions:

\begin{itemize}
\item We explored the use of ChatGPT in the domain of open-ended question grading and feedback.
\item We assessed the responses of experts (2 people) and non-experts  (40 people) using ChatGPT.
\item We compared the grading provided by ChatGPT using a typical metric, identifying and explaining any potential inconsistencies.
\end{itemize}

%\gnote{colocar aqui duas RQs para as respostas}

\section{Why ChatGPT?}

%O avanço tecnológico trouxe uma enorme quantidade de dados ao mundo, incluindo dados estruturados (ex., tabelas) e desestruturados, como imagens, sons e textos.  O poder computacional também melhorou significativamente, com um aumento de 10 vezes na capacidade de processamento e memória das GPUs nos últimos quatro anos. 

The recent technological advancements, such as the significant improvement in computational power and the enormous amount of data stored in structured and unstructured formats, have greatly benefited the field of Machine Learning (\textit{ML}), especially Deep Learning (\textit{DL}). DL, in particular, has revolutionized various domains of knowledge, such as image and speech recognition~\cite{hernandez2019systematic}.

The early DL models for long sequences (such as texts) processed inputs sequentially~\cite{li2020survey}, which required larger models, more time, and computational power for training. Additionally, these models struggled to relate different parts of the sequences, resulting in limitations in learning~\cite{ma2019universal}. However, with the introduction of \textit{Transformers} models, parallel processing of long-term sequences became possible, enhancing the learning capacity and accelerating the process~\cite{vaswani2017attention}. These improvements culminated in \emph{Large Language Models} (LLMs), which provided significant advances in text processing and natural language understanding~\cite{zhao2023survey}.

\vspace{0.2cm}
\noindent
\textbf{Transformers Models.} Transformers models are a widely used deep neural network architecture in Natural Language Processing (\textit{NLP}). Unlike previous neural network models that processed elements sequentially, Transformers introduced the attention mechanism~\cite{vaswani2017attention}.

The attention mechanism allows the model to assign different weights to specific parts of the input during training. Instead of solely relying on sequential order, the model can focus on parts of the sequence that are more relevant to the task at hand, capturing complex and long-range dependencies. For instance, when given a paragraph as input, while previous models "read" the paragraph sequentially, the attention mechanism allows Transformers to assign higher importance to specific words. This way, the model can capture long-range relationships between words, even when there are several words between them.

Moreover, the attention mechanism enables Transformers to process data in parallel, dividing the input sequence into multiple parts and performing attention operations independently. This leads to efficient and simultaneous processing of information, resulting in smaller models compared to traditional ones. This characteristic is one of the main reasons why these models can efficiently and scalably process and generate text on a large scale~\cite{vaswani2017attention,zhao2023survey}.

\vspace{0.2cm}
\noindent
\textbf{Large Language Models.} LLMs are Transformers models with millions, billions, or even more parameters, extensively trained on large textual datasets, such as libraries of books, web articles, and conversations on social networks. Through this diverse training data, these models acquire a deep understanding of the structure, grammar, and semantic context of human language.

Previously, training and utilizing these models for specific tasks required considerable computational resources and advanced technical knowledge to develop, deploy, and enhance the models. However, platforms like ChatGPT and similar ones have democratized access to LLMs, enabling individuals without ML expertise to interact intuitively with these models through chat interfaces, such as virtual assistants.

This mode of interaction has brought about an alternative learning paradigm in the field of ML: prompt-based learning. Instead of refining the model traditionally, through providing more data and adjusting parameters, tasks are reformulated as textual prompts. An appropriate prompt can shape the model's behavior, directing the desired output without the need for conventional fine-tuning~\cite{liu2023pretrain}. As a result, LLMs demonstrate impressive ability to perform complex tasks, even when trained with few examples (few-shot learning~\cite{brown2020language}) or no examples (zero-shot learning~\cite{kojima2022large}). This qualifies them for a wide range of NLP activities, including automatic translation, text generation, and document summarization~\cite{zhao2023survey}.

\vspace{0.2cm}
\noindent
\textbf{ChatGPT in Education.} ChatGPT is a specific implementation of an LLM based on the GPT (\emph{Generative Pre-trained Transformer}) architecture~\cite{brown2020language}. This tool has the potential to promote improvements in learning and teaching experiences at various levels, from school education to university and professional development.

One of the advantages of ChatGPT, along with other LLMs, is the ability to offer personalized learning, taking into account the preferences, abilities, and individual needs of each student. This personalization can contribute to making the learning experience more effective and engaging~\cite{kasneci2023chatgpt}.

%Para o presente trabalho, estamos interessados em investigar o uso do ChatGPT como estratégia complementar de correção e \emph{feedback} de questões abertas.

%\section{Method}

%\gnote{Lucena, vc consegue revisar essa seção, colocando referencias e fazendo apontamento de melhorias?}

%\lucena{Ajustando aqui}

%% versão anterior
%% Para conduzir esse estudo, utilizamos de questionários como mecanismo de coleta de respostas para as questões abertas. O processo de desenho dos questionários está detalhado na Seção~\ref{sec:questionarios}. Em seguida, para avaliação das respostas dos profissionais, utilizamos técnicas de engenharia de prompt, que estão descritas na Seção~\ref{sec:prompt}.

%Section \ref{sec:qp} presents the research questions (RQs) that motivate our work. To carry out this study, we used questionnaires as a tool to collect responses to open-ended questions. The questionnaire design process is detailed in Section~\ref{sec:questionarios}. Subsequently, to evaluate the professionals' responses, we employed prompt engineering techniques, described in Section~\ref{sec:prompt}. 

\section{Research Questions}\label{sec:qp}

This work aims to provide answers to two research questions:

\begin{itemize}
\item[\textbf{RQ1:}] Considering the responses of experts, what is the quality of the grading provided by ChatGPT?
\item[\textbf{RQ2:}] Considering the responses of non-experts, what is the quality of the grading provided by ChatGPT?
\end{itemize}

To answer \textbf{RQ1}, we asked experts to respond to six open-ended questions. These questions were corrected using ChatGPT. For comparison purposes, we also asked ChatGPT to answer the same questions; and we also corrected the questions answered by ChatGPT with ChatGPT.
After refining the responses to the six questions, to answer \textbf{RQ2}, we asked a larger group of developers, but without expertise in that specific domain, to respond to the questions.

\section{Method: Questionnaire}\label{sec:questionarios}

%% versão anterior
%% Nessa seção descrevermos o processo de design, piloto e deploy do questionário.

In this section, we will present in detail the process of creating, testing, and implementing the questionnaire. We will discuss the steps involved in formulating the questions, as well as the methods used to ensure the validity and reliability of the results. Additionally, we will describe the process of testing the questionnaire with a pilot group and how the results of this test were used to improve the final experiment.

%% versão anterior
%% \textbf{Design do questionário.} Para elaborar o questionário, escolhemos dois temas em que contamos com especialistas no time de Educação da \zup. Essa decisão foi tomada pois assim poderíamos gerar \emph{prompts} que comparassem as respostas dos especialistas com as respostas dos participantes do experimento, assim, diminuindo o viés de confirmação no cálculo de similaridade realizado pelo ChatGPT \gnote{é vies de confirmação ou é outro nome?}. Para seleção das questões, escolhemos três questões de cada tema. Decidimos por um pequeno número de questões abertas como forma de não atolar\gnote{ver um verbo melhor} os participantes da pesquisa, uma vez que estes utilizavam de seu horário de trabalho para responder as questões. As questões foram organizadas em progressivo grau de dificuldade, da mais fácil a mais difícil. As questões escolhidas (Q\emph{n}), junto com as respostas dos especialistas (R\emph{n}), foram:

%\lucena{reescrevi esse trecho}

\subsection{Questionnaire design} 
To create the questionnaire, we selected two topics for which we had experts available in the Education team at \zup: (1) caching and (2) stress and performance testing . This decision was made so that we could generate prompts that compared the responses of the experts with the responses of the participants in the experiment. This way, we could reduce confirmation bias in the similarity calculation performed by ChatGPT~\cite{witteveen2019paraphrasing}. For the selection of questions, we chose three questions for each topic. We opted for a small number of open-ended questions to avoid overburdening the research participants, as they were using their work hours to answer the questions. The questions were organized in ascending order of difficulty, starting with the easiest and ending with the most difficult. The chosen questions (Q\emph{n}), along with the responses of the experts (R\emph{n}), were:

\vspace{0.2cm}
\noindent
\textbf{\emph{Caching}}

\begin{enumerate}
    \item[(Q1)] Explain in your own words what you understand about \emph{caching} in REST applications.
    \item[(R1)] ``\emph{Caching} is a technique that allows storing frequently accessed data in memory to reduce the complexity cost of querying them. Nowadays, there are several ways to apply \emph{caching} in REST APIs, starting on the server-side through techniques of local and distributed caching. It is also possible to enable caching on the client-side, where through the use of HTTP protocol headers, policies are defined to govern the \emph{caching} behavior, such as using versions, expiration time, and also specifying which clients can store the data, referring to the end user's browser and/or CDNs.''

    \item[(Q2)] Explain briefly how the two types of \emph{caching} work: client-side caching and server-side caching.
    \item[(R2)] ``\emph{Server-side caching can be served locally, causing a portion of the server's memory heap to be used for storing data that has high network or computation cost and is frequently accessed. This strategy should be used in scenarios where the system architecture is monolithic or there is a restriction that only one instance of the system is used. Another way to provide caching at the application layer is through the use of distributed cache providers, favoring a global point of access that is shared among instances, facilitating data synchronization with the source of truth.}''

    \item[(Q3)] Explain cache invalidation in REST applications and present a way to address it.
    \item[(R3)] ``\emph{Cache invalidation is an operation that aims to keep the caching lean and consistent. To provide these guarantees, invalidation policies must be used. Some examples are Least Recently Used (LRU), which aims to remove from the cache the data that has not been accessed recently. Another example of a policy is Least Frequently Used (LFU), which aims to remove from the cache the data that is least accessed. There are also providers that work with expiration policies, where the data enters with a duration time, and upon reaching a certain time, they are automatically removed from the cache.}''

\end{enumerate}

\noindent
\textbf{\emph{Stress and performance testing}}

\begin{enumerate}
    \item[(Q4)] Explain the concept of load and stress testing.
    \item[(R4)] ``\emph{Load testing means verifying how an application or system behaves under an expected workload, which can be small, moderate, or large. Additionally, this workload is applied for a certain interval of time, such as minutes or hours, to validate the system's stability and detect possible problems in resource usage, such as memory, CPU, disk, or connections to a database, for example. It is important to understand that load testing does not exceed the expected or designed capacity for an application or system. On the other hand, stress testing is related to verifying how an application or system behaves when subjected to a very high and intense workload, usually a workload higher than expected or specified in the requirements. The idea here is to subject the application beyond its designed capacity in order to detect problems or bottlenecks in resource or internal component usage. The goal is to discover how the system behaves under extreme pressure, such as traffic spikes, excessive resource usage, hardware failures, or abnormal conditions.}''

    \item[(Q5)] What are the main metrics used to evaluate the performance of an application during a load test?
    \item[(R5)] ``\emph{Generally, for a web application, including REST APIs, the main metrics we collect and evaluate are: response time, throughput (number of operations per unit of time), and error rate. There is a well-known method called the `RED Method,' which basically recommends evaluating these 3 metrics for request-based services. For non-request-based applications, such as batch processing or streaming services, other metrics like CPU, memory, or network are also usually collected and evaluated.}''

    \item[(Q6)] What are the best practices for conducting load tests on applications that expose REST APIs?
    \item[(R6)] ``\emph{There are several important practices when conducting load tests, such as defining the use cases to be validated and the expectations of the expected workload. It is also important to define which metrics are relevant for the test, as they will help identify performance issues and bottlenecks (here, the `RED Method' can be adopted). Another point is to run load tests against an application in production or a similar production-like environment, such as a staging environment; this way, we will obtain numbers close to the reality of the system. And last but not least, applying the tests with a realistic data set whenever possible.}''

\end{enumerate}

%% versao anterior
%% O questionário foi implementado por meio da plataforma TypeForm. 
%% O questionário era anônimo, e todas as perguntas eram obrigatórias. No entanto, por conta da implementação de políticas de privacidade na empresa, o questionário não contava com questões demográficas; somente as seis questões abertas. Antes de cada grupo de perguntas, introduzimos o objetivo do trabalho, além dos conceito \emph{caching} e de testes de estresse e desempenho.

The questionnaire used in the study was implemented through the TypeForm platform. It was designed to be anonymous, and all questions were mandatory. However, due to the implementation of privacy policies at \zup, the questionnaire did not include demographic questions. Instead, it contained only six open-ended questions. Before each group of questions, we presented the purpose of the study and introduced the concepts of caching and stress and performance testing.
%\lucena{Gustavo, aqui precisa colocar a url do site do typeform no rodapé?}

\subsection{Questionnaire Pilot} 
Before making the survey available and collecting responses from a larger population, we first conducted a pilot with two objectives: 1) to evaluate the understanding of the presented questions, and 2) to assess the quality of the prompts used for question grading in ChatGPT.

To conduct the pilot, we sent an invitation message to a virtual space with developers from \zup who discuss topics related to caching and to stress and performance testing\footnote{At \zup, the use of such spaces for forming discussion groups on relevant topics is common.}. In the message, we introduced the objectives of the work along with the link to the online form. In total, six participants responded to the questions. According to TypeForm statistics, 65 people opened the form, while 49 started to answer it. Out of those, only six completed the responses to all questions. 
The average time for completing the activity was 38 minutes. With the first set of responses, we initiated the process of engineering the prompt for grading (further details in Section~\ref{sec:prompt}). After the gradings were completed, we shared the responses together with the evaluation from ChatGPT with the same virtual space where we invited the individuals, as a way to request feedback on the evaluation performed by ChatGPT. Two people provided feedback, which helped us improve the provided prompt (e.g., ``All my responses followed the same feedback pattern. It presents a correct and general approach on the topic, mentioning some important points. However, \textbf{it could provide more details and real-world examples}. Sometimes, it asks for more technical details, and in others, it requests clear explanations for all audiences.'').

\subsection{Actual Questionnaire} 
After conducting the pilot and refining the prompts based on participants' feedback, we sent the questionnaire to a larger population of developers at \zup. Initially, we selected 100 individuals who had already participated in at least one of the training sessions offered by the Education team. We made this decision to target the questions towards developers who are more accustomed to study activities during their work environment. Out of the 100 participants, we selected 50 who had successfully completed at least one training, meaning they achieved the minimum required average to pass the course; similarly, we selected 50 who had not completed any training. Subsequently, we individually invited each participant through a message sent via the private enterprise chat platform. Similar to the pilot, in the message we expressed our interest and the study's objective, along with the link to the online questionnaire. As in the pilot, the questionnaire was anonymous and no questions regarding demographic information were asked. Over the course of a week, we sent reminder messages to those who had not indicated whether they had responded to the questionnaire or not. According to TypeForm statistics, 90 individuals (53 from the  group who completed at leas one training and 37 from the group who did not) opened the form, while 76 (43+33) started to respond to it. In the end, we obtained 40 (\totalPessoasCompletaramTreinamentos+\totalPessoasNaoCompletaramTreinamentos) responses. The average time for completing the activity was 42 minutes.

\subsection{Prompt Engineering}\label{sec:prompt}

%\gnote{Lucena, consegue enriquecer com referencias aqui?}

%\lucena{Adicionar algumas referências que acho relevantes no contexto}

The term \emph{prompt} refers to a set of instructions provided to a LLM (Language Model) as a way to customize or refine its capabilities. The prompt defines the context of the conversation, informs the LLM about what information is relevant, and specifies the type of output expected. The quality of the output generated by an LLM is directly related to the quality of the prompt provided by the user.

Prompt engineering~\cite{liu2022design}, on the other hand, refers to the process of adjusting or improving the responses generated by LLMs~\cite{zhao2023survey}. In other words, prompt engineering involves training LLMs via prompts. The process of prompt engineering involves modifying the prompts to obtain more desirable results. This fine-tuning technique optimizes human-computer interaction by refining text generation based on the users' needs and expectations.

\subsubsection{Prompt Engineering Phases}\label{sec:prompt-engineering}

We describe our prompt engineering steps below.

\vspace{0.2cm}
\noindent
\textbf{Prompt V1.} In the first prompt version, we used ChatGPT as an oracle for open-ended questions. Additionally, we included technical aspects that, although relevant to the company's context, were not necessary for answering the open-ended questions. The initial prompt is as follows:

\MyBox{   
Suppose you are an expert in creating web applications using REST APIs.\\
In addition to being an expert, you grade exams on this topic.\\
Consider the following question and answer:\\
Q: Explain in your own words what you understand about caching in REST applications?\\
R: \{\}\\
What score would you give to this answer on a scale from 0 to 10?\\
Return the answer in a JSON format, with a variable 'score' for your rating, and another variable 'explanation' for the justification of this score.\\
Your explanation should have at least 20 words. In the explanation, identify any knowledge gaps and explain how to minimize them using real-world examples.\\
If no response is provided, indicate 'No response was given,' and assign a score of zero.
}

In the prompt above, we illustrated a question about \emph{caching}. We used the approach of creating a prompt for each question, only changing the context (first paragraph) and the question itself.
The response is provided to the prompt in a way that can be parameterized (indicated by \texttt{\{\}} in the prompt).
However, when evaluating this first prompt on the pilot data, we observed that the scores given by the model were consistently high, even for simple answers, indicating a possible model miscalibration.

\vspace{0.2cm}
\noindent
\textbf{Prompt V2.} 
As an attempt to increase the model's evaluation rigor, we removed the first sentence from the prompt and replaced it with the following instruction: ``I need you to grade the exams using the highest grading standard you can.'' The rest of the prompt remained unchanged.
Furthermore, the first prompt was designed to be specific to the context of programming in Java. However, it was observed that the questions were theoretical and did not require knowledge of a specific programming language. Therefore, this information was also removed.
With these modifications, it was possible to observe an increase in the evaluation rigor of the questions.

\vspace{0.2cm}
\noindent
\textbf{Prompt V3.} 
Although the V2 version of the prompt increased the rigor of question evaluation, it still has an important limitation: relying solely on the knowledge base and interpretation of ChatGPT, which is known to provide incorrect information~\cite{ray2023chatgpt}. In order to minimize this limitation, we designed a new prompt by providing the expert's answer and then we asked to compare the student's answer with the expert's answer.

\vspace{0.2cm}
\noindent
\textbf{Prompt V4.} After the first round of pilot evaluation, the scores with the explanations provided by ChatGPT were shared with the pilot participants. Among the feedback received, it was suggested that the explanation would be more illustrative if it included real-world examples. Therefore, we added a final instruction at the end of the prompt, indicating that ``Whenever possible, the explanation should include real-world examples.'' The final prompt generated and used for the correction of the remaining open-ended questions can be seen below:

\MyBox{I need you to grade the exams using the highest grading standards possible.\

Consider the following question and the answer provided by an expert:\

Q: Explain in your own words what you understand about \emph{caching} in REST applications?\

R Expert: \emph{Caching} is a technique that allows storing data frequently accessed in memory to reduce the complexity cost of querying it. Nowadays, there are various ways to apply \emph{caching} in REST APIs, starting on the server-side through local and distributed cache techniques. It is also possible to enable client-side \emph{caching}, where HTTP protocol headers define policies that govern \emph{caching} behavior, such as versioning, expiration time, and defining which clients can store the data, referring to the end-user's browser and/or CDNs.\

Now, consider the student's response provided below. \

R Student: \{\}\

What grade would you give to the student's response, considering the expert's answer, on a scale from 0 to 10?\

Return the response in a JSON format, with a variable 'grade', containing your grade, and another variable 'explanation' with the explanation for this grade.\

Your explanation must have at least 20 words. In the explanation, identify knowledge gaps and explain how to minimize these gaps using real-world examples.\

If no response is provided, inform 'No response provided', and give a grade of zero.}

\subsubsection{Prompts Execution}

In this article, we utilized ChatGPT through its API\footnote{\url{https://platform.openai.com/docs/api-reference}}. We used the GPT-4 model with the following configurations: a maximum of 100 tokens and a temperature of 0.6; the temperature is a hyperparameter that affects the probability of token distributions in the model. It is possible to adjust the temperature to control the diversity and creativity of the generated texts. The temperature value ranges from 0 to 1. While a high temperature (e.g., 0.7+) leads to more diverse results, a low temperature (e.g., 0.2) tends to produce more deterministic results.

The value of 0.6 was chosen to obtain slightly different corrections for similar questions. The maximum token value was set to make the GPT's corrections more concise and direct.

\subsection{Quality Metrics}\label{sec:metrica}

Like other LLMs, ChatGPT has a well-known limitation regarding false or distorted perceptions of information generated by the model itself. These ``hallucinations'' (a technical term used to describe this limitation) can cause the LLM to generate answers that appear correct when, in fact, they are not.

To minimize this limitation and complement the grading provided by ChatGPT, we also calculated the cosine similarity metric between the response provided by the expert and the response provided by the study participant.
The cosine similarity metric is widely used to measure similarity between vectors in multidimensional spaces~\cite{li2013distance, lahitani2016cosine, singh2021text}.
By calculating the cosine similarity between the participants' responses and a reference standard (expert's response), we can obtain an objective measure of how close the responses are to the desired pattern.

To calculate this metric, we used the implementation available in the sbert library\footnote{\url{https://www.sbert.net/}}. In general, to calculate cosine similarity, we first convert the responses into numerical vectors.
Then, we multiply these vectors with each other and sum the results. Finally, we divide the multiplication result by the product of the sizes of the vectors.
The final value obtained ranges from -1 to 1, where 1 indicates that the responses are very similar, while -1 indicates that the responses are very different. In summary, this metric measures the projection of the participant's response onto the direction of the expert's response, providing a measure of their similarity.

\section{Results}

We organize the results in terms of the research questions. First, to answer \textbf{RQ1}, we present an analysis of the correctness of experts' responses (Section~\ref{sec:respostas-specs}), comparing them to ChatGPT's responses. Next, to address \textbf{RQ2}, we present an analysis of the correctness of students' responses (Section~\ref{sec:respostas-alunos}).

\subsection{Grading experts' responses}\label{sec:respostas-specs}

We started by evaluating the responses provided by experts (as presented in Section~\ref{sec:questionarios}) using ChatGPT. To conduct this evaluation, we asked ChatGPT to grade the responses given by the experts. To do this, we had to edit the \emph{prompt} to remove the lines indicating the expert's evaluation. As a result, the expert's response was corrected as if it were a student's answer. The scores assigned by ChatGPT to the experts' responses are presented in the Experts'' column of Table~\ref{tab:notas-specs}. After collecting and grading the experts' responses with ChatGPT, we asked ChatGPT to answer the six open questions. To do this, we used the following prompt: ``\texttt{Using the highest grading scale you can, provide an answer of up to 100 words for the following question: QN}''. The limit of 100 words was established since the average word count in the experts' responses was 114 words.
These responses provided by ChatGPT were then also corrected by ChatGPT itself. The result of this correction is available in the ``ChatGPT'' column of Table~\ref{tab:notas-specs}.

\begin{table}[h!]
    \centering
    \small
    \caption{
    Evaluation of experts' and ChatGPT's responses, graded by ChatGPT. The column \# Words'' presents the size of the response provided by the expert (or ChatGPT), measured by the number of words. The column ``Cos Sim'' presents the result of the cosine similarity metric.}  
    \label{tab:notas-specs}
    \begin{tabular}{c|cc|cc|c}
    \toprule
    & \multicolumn{2}{c|}{\textbf{Experts}} & \multicolumn{2}{c|}{\textbf{ChatGPT}} \\
    \textbf{Question} & \textbf{\# Words} & \textbf{Grade} & \textbf{\# Words} & \textbf{Grade} & \textbf{Cos Sim}  \\
    \midrule
    Q1  &  107  & 8    & 94  & 9   & 0.8410 \\
    Q2  &  109  & 7    & 105 & 9   & 0.7338 \\
    Q3  &  94   & 7    & 82  & 8.5 & 0.7492 \\
    Q4  &  178  & 8.5  & 93  & 9   & 0.8546 \\
    Q5  &  88   & 9    & 61  & 9   & 0.5429 \\
    Q6  &  110  & 8.5  & 93  & 9   & 0.8046 \\
    \bottomrule
    \end{tabular}
\end{table}

%\gnote{será que vale a pena perguntar pro gpt qual seria a resposta ideal das 6 perguntas?}
%\gnote{qual é a similaridade desse pra outro?}

\vspace{0.2cm}
\noindent
\textbf{Grading of experts' responses by ChatGPT.} As one could see, the answers to the first group of questions about caching (Q1, Q2, and Q3) consistently received lower scores compared to the second group. As highlighted in Section~\ref{sec:questionarios}, the responses related to caching were provided by one expert, while the responses concerning stress testing and performance were provided by another expert.

Upon analyzing the feedback provided by ChatGPT, we observed comments that might have influenced the received scores. For instance, when considering the first set of answers about caching, ChatGPT indicated that the responses presented basic concepts but did not adequately explain or provide examples. In Q2, which received a score of 7 and asked about the difference between client-side and server-side caching, the expert explained server-side caching but did not mention the second approach, client-side caching. ChatGPT identified this gap and highlighted it in its correction.
``\emph{The student explained server-side caching well, including monolithic and distributed strategies, but did not mention client-side caching. In client-side caching, data is stored on the user's device, such as browsers and APIs, speeding up page loading. For example, saving images and CSS from the site of the last visit.}''

Another interesting example to highlight is the explanation provided for Q5, which inquired about the main load testing metrics. In this explanation, ChatGPT indicated that: ``\emph{The student's response correctly addressed the response time, throughput, and error rate metrics, mentioned the RED Method, and cited examples of other situations. However, \textbf{it could have explicitly explained the relationship of these metrics} with the user experience and mentioned more specific metrics for other applications, such as latency indexes for real-time systems}''. In other words, ChatGPT correctly identified the items provided in the responses but did not recognize a potential relationship among these items---even though this relationship was not questioned in the prompt.

\vspace{0.2cm}
\noindent
\textbf{Perception of ChatGPT's grading by the experts.}
Next, we sent the corrections provided by ChatGPT to the experts who wrote the responses. When we inquired whether the experts agreed with the evaluations, they generally concurred with the assessments. In particular, regarding question Q3, where there was a higher disagreement between the experts' and ChatGPT's responses, the expert who provided the answer commented that: ``\emph{[in this question] he was precise and noticed that my response was incomplete. I believe the score was higher than it should have been. I would give myself a 5.}'' However, for Q4, a point of disagreement was observed as the expert mentioned that ``\emph{This explanation doesn't make much sense since what she says is missing from the response is actually there.}'' referring to the difference between load testing and performance testing.

\vspace{0.2cm}
\noindent
\textbf{Corrections of ChatGPT's responses by ChatGPT.} The gradings provided for the expert responses questioned whether ChatGPT could provide even more elaborate answers. The column labeled ``ChatGPT'' shows the result of this investigation. As evident, in general, the responses provided by ChatGPT received higher scores when compared to the responses provided by the experts. Particularly, the difference was greater in the group of questions about \emph{caching}: 2 points of difference for question Q2 and 1.5 points of difference for question Q3. For example, ChatGPT's response to Q2 addressed both server-side and client-side caching concepts: ``\emph{Client-side and server-side caching are strategies to temporarily store data to improve performance and loading speed. In client-side caching, browsers store information locally on the user's device, such as HTML, CSS, and JavaScript files. This reduces the need to repeatedly retrieve resources from the server, saving loading time and bandwidth. On the other hand, server-side caching involves storing information, such as request responses, directly on the server. In these cases, if the information is already cached, server responses are faster, optimizing the user experience.}'' However, for the second group of questions about stress and performance testing, a smaller variation was observed between the expert's grading and ChatGPT's grading: only 0.5 points for Q4 and also 0.5 points for Q6.

\vspace{0.2cm}
\noindent
\textbf{Comparison of expert and ChatGPT gradings.} Although the gradings provided by ChatGPT for the expert responses offer an initial means of comparison, as discussed in Section~\ref{sec:metrica}, relying solely on ChatGPT's corrections may be insufficient, as it can exhibit hallucinations, meaning it provides false or distorted insights generated by the model itself~\cite{ji2023survey}. To complement ChatGPT's corrections, we employ the Cosine Similarity metric, presented in Table~\ref{tab:notas-specs}, under the column labeled ``Cos Sim.'' When closer to 1, there is a higher likelihood that the responses provided by the experts and ChatGPT are similar. As evident, in general, the responses exhibited a high degree of similarity ($>$0.7). However, only response Q5 showed a lower similarity value, even though both responses received a score of 9, according to ChatGPT's correction. Upon evaluating these two responses, it was observed that the expert and ChatGPT approached slightly different topics, although both were correct. For instance, in Q5, the expert commented: ``\emph{Usually, in a web application, including REST APIs, the main metrics we collect and evaluate are: response time, throughput (number of operations per unit of time), and error rate. In fact, there is a well-known method called the 'RED Method,' which basically recommends evaluating these 3 metrics for request-based services and applications.}'' On the other hand, ChatGPT's response was more generic and did not mention the existence of the 'Red Method': ``\emph{The main metrics used to evaluate the performance of applications during a load test include response time, throughput, the number of simultaneous users, resource utilization (CPU, memory, disk, and network), and system errors or failures. These metrics help identify bottlenecks, determine scalability, and ensure system stability under different load conditions.}''

\subsection{Grading students' responses}\label{sec:respostas-alunos}

After the individual evaluation of expert responses, we proceeded with the evaluation of the study participants' responses. As highlighted in Section~\ref{sec:prompt-engineering}, in the final version of the prompt, we instructed ChatGPT to compare the student's response with the expert's response. This approach aimed to minimize known hallucination biases often observed in LLMs like ChatGPT~\cite{ji2023survey,ray2023chatgpt}.
Furthermore, based on the feedback from the experts, the initially provided responses were adjusted by the authors of this article to incorporate the insights highlighted by ChatGPT.
The following Table~\ref{tab:notas-alunos} provides a summary of the evaluations from the study participants.

\vspace{0.2cm}
\noindent
\textbf{Difference in scores between participant groups.} Although the participants of the survey did not undergo any training on caching or stress and load testing, our initial hypothesis was that participants who had completed at least one technical training would have less difficulty answering the questions (indicated by higher scores) compared to participants who had not completed any technical training.
In particular, for questions Q1 and Q2, it was possible to observe that participants who completed at least one training scored at least one point higher than those who did not complete any training (for Q1, median of 6 for the completed group, while median of 5 for the non-completed group; for Q2, the completed group scored 7 at the median, while the non-completed group scored 6).
However, for questions Q4 and Q5, the difference between the groups was small (around 0.29 points in Q4 and around 0.1 points in Q5, on average). On the other hand, for questions Q3 and Q6, participants who did not complete any training received higher scores compared to those who completed a training (for Q3, median of 5 points for the completed group and 7 points for the non-completed group, while for Q6, the average was again 5 points for the completed group and 7 points for the non-completed group).

\begin{table}[h!]
    \centering
    \caption{Assessments of study participants' responses.} 
    \label{tab:notas-alunos}
    \begin{tabular}{ll|ccc}
    \toprule
    \textbf{Question} & \textbf{Group} & \textbf{Avg.} & \textbf{Median} & \textbf{Std. Dev.} \\
    \midrule
    \multirow{2}{*}{Q1}  & Completed     & 5,50 & 6 & 1,71 \\
                         & Non-Completed & 4,96 & 5 & 0,92 \\
    \multirow{2}{*}{Q2}  & Completed     & 7,29 & 7 & 0,47 \\
                         & Non-Completed & 5,65 & 6 & 0,69 \\
    \multirow{2}{*}{Q3}  & Completed     & 5,21 & 5 & 1,86 \\
                         & Non-Completed & 6,04 & 7 & 1,94 \\
    \multirow{2}{*}{Q4}  & Completed     & 5,29 & 6 & 2,26 \\
                         & Non-Completed & 5,00 & 5 & 2,72 \\
    \multirow{2}{*}{Q5}  & Completed     & 5,21 & 6 & 2,82 \\
                         & Non-Completed & 5,31 & 6 & 2,66 \\
    \multirow{2}{*}{Q6}  & Completed     & 5,21 & 5 & 1,86 \\
                         & Non-Completed & 6,04 & 7 & 1,94 \\
    \bottomrule
    \end{tabular}
\end{table}

\begin{figure*}[thp]
\begin{center}
$
\begin{array}{ccc}
  \textsc{Q1} & \textsc{Q2} & \textsc{Q3}\\
  \includegraphics[width=.32\textwidth]{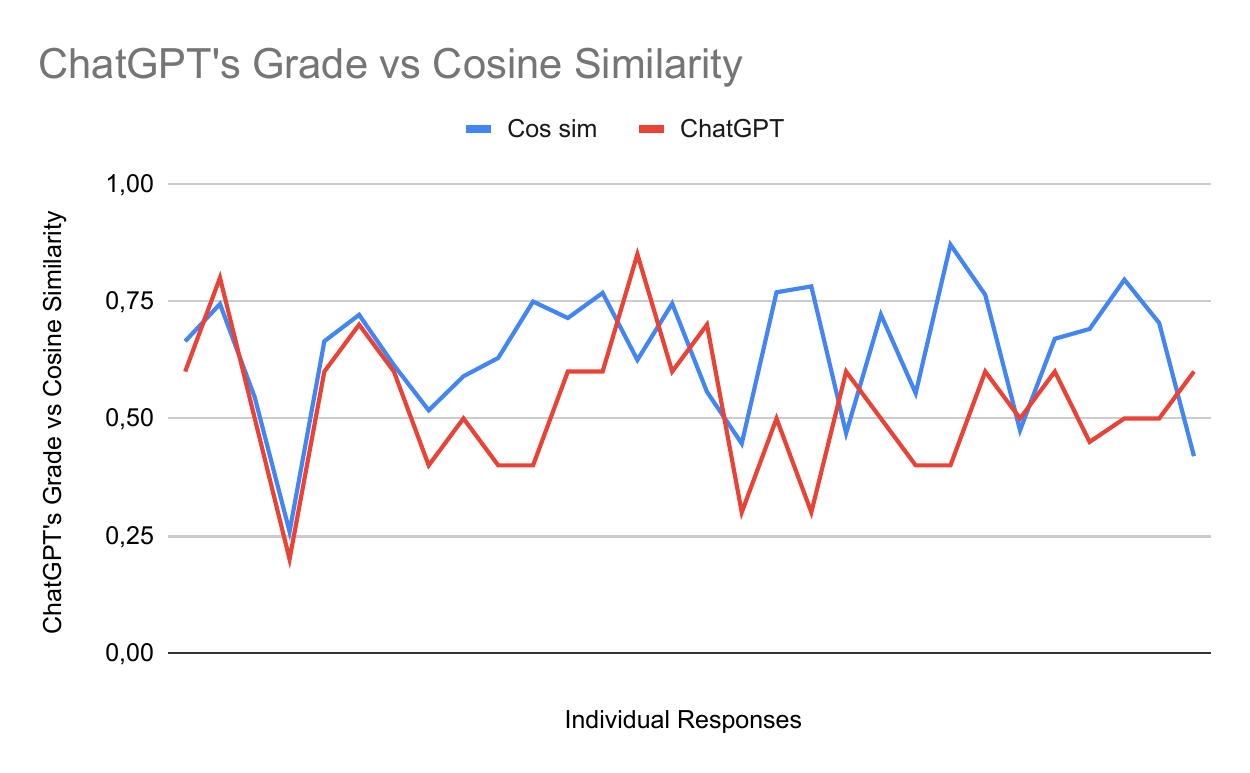}&
  \includegraphics[width=.32\textwidth]{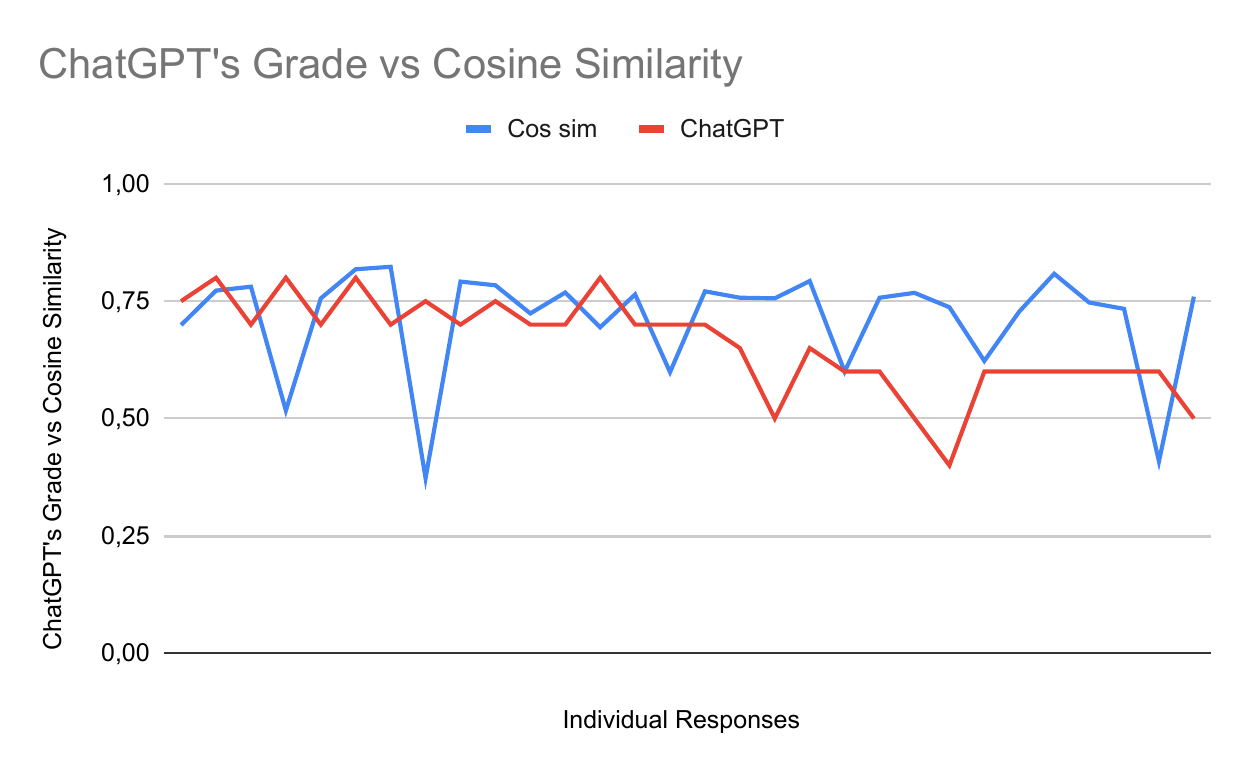}&
  \includegraphics[width=.32\textwidth]{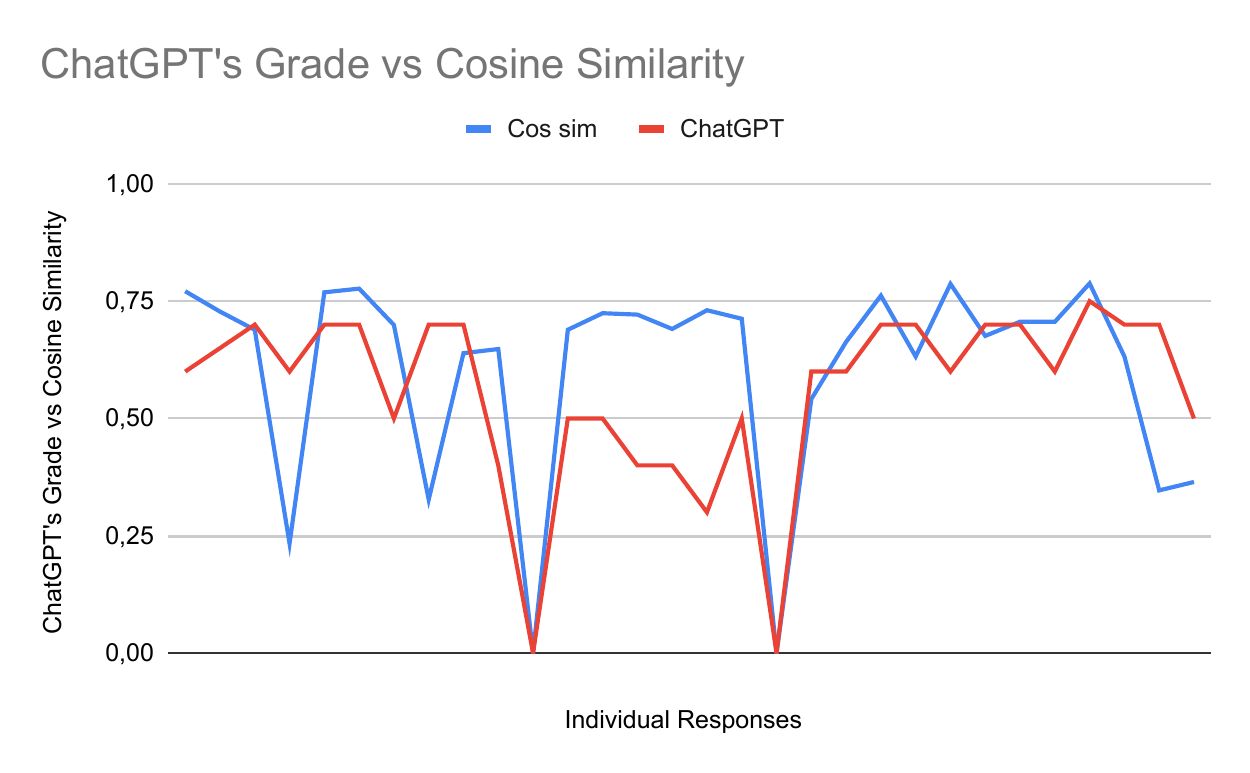}\\
  \textsc{Q4} & \textsc{Q5} & \textsc{Q6}\\
  \includegraphics[width=.32\textwidth]{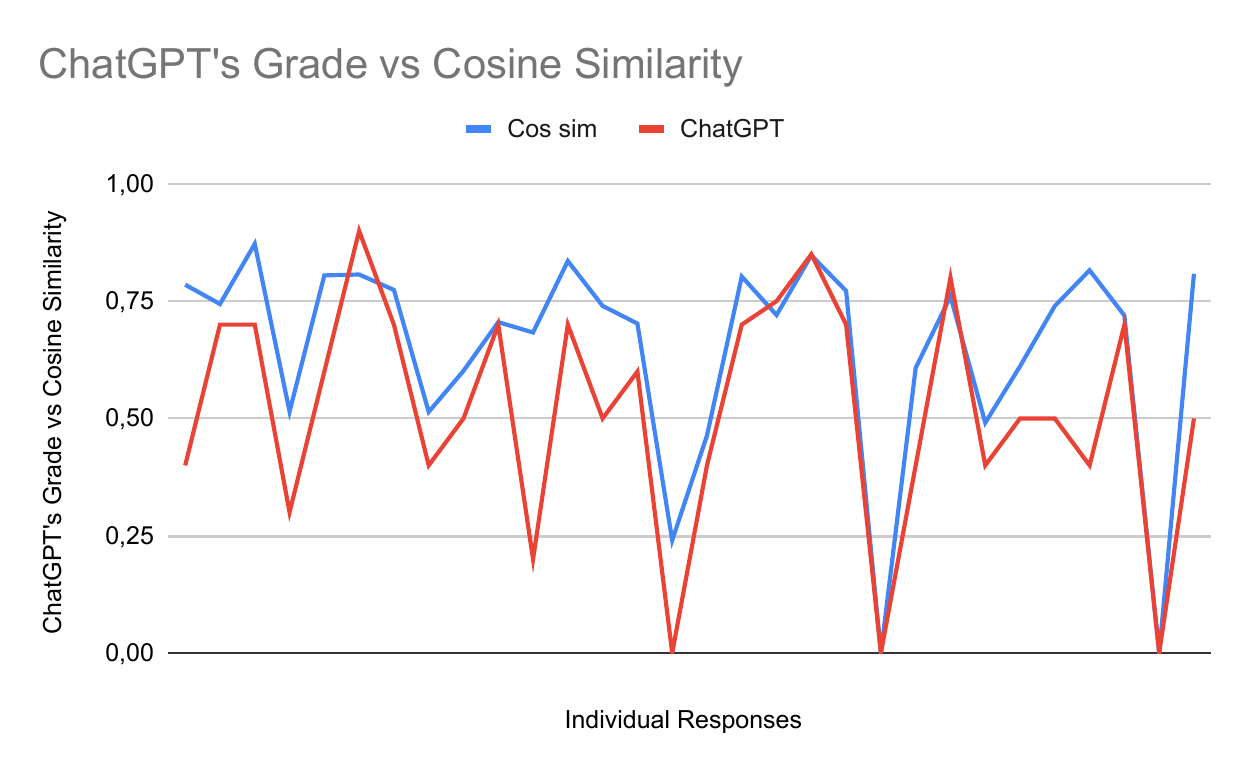}&
  \includegraphics[width=.32\textwidth]{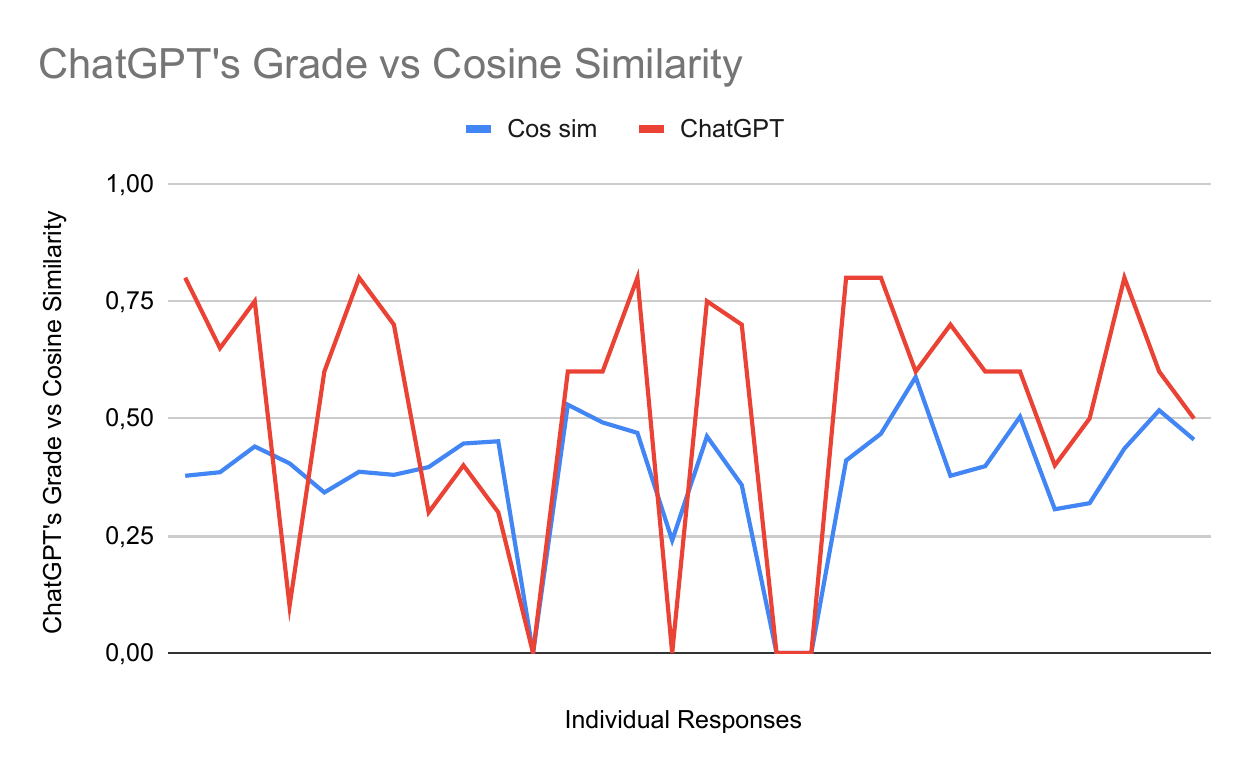}&
  \includegraphics[width=.32\textwidth]{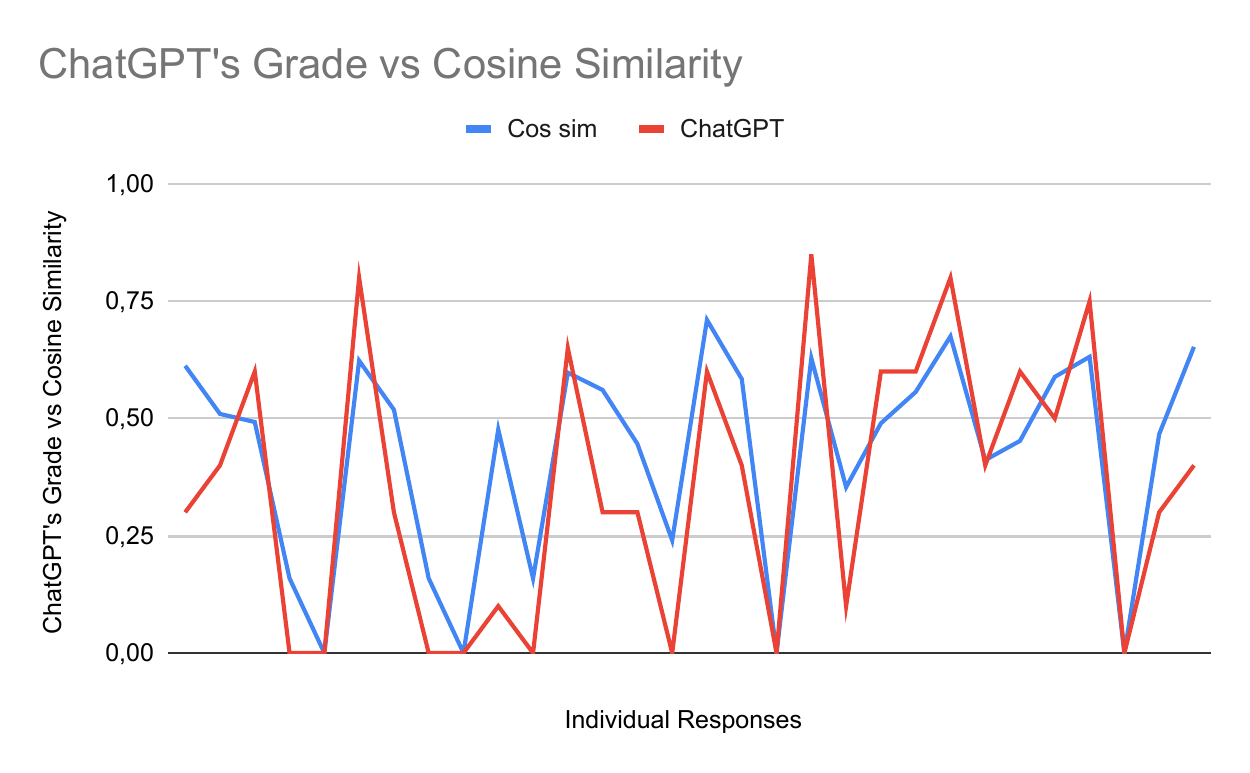}\\
\end{array}
$
\end{center}
\caption{
Comparison of responses provided by ChatGPT and those obtained by calculating the cosine similarity metric. The red line indicates the question grade value indicated by ChatGPT, while the blue line indicates the resulting metric value.}
\label{fig:respostas}
\end{figure*}

\begin{figure*}[thp]
\begin{center}
$
\begin{array}{ccc}
  \textsc{Q1} & \textsc{Q2} & \textsc{Q3}\\
  \includegraphics[width=.30\textwidth, clip=true, trim= 30px 40px 30px 50px]{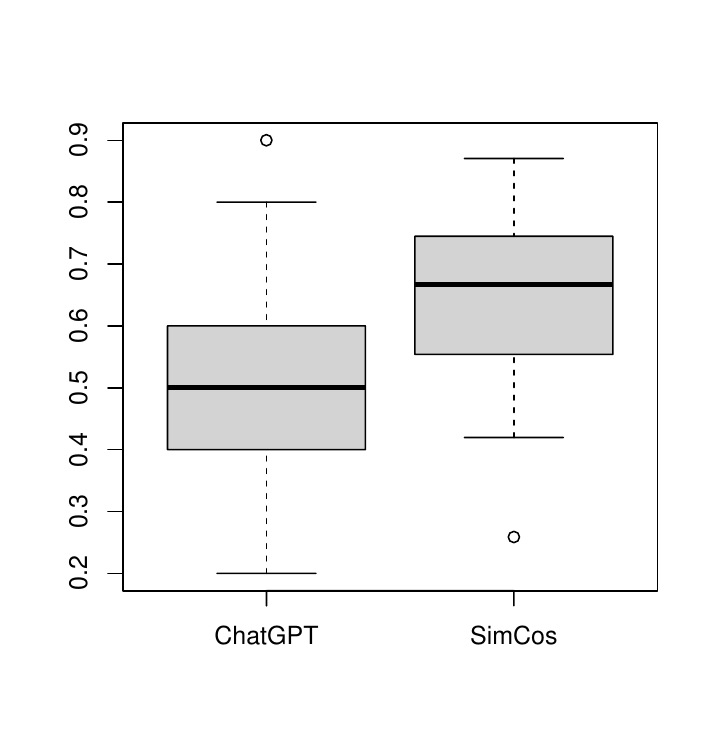}&
  \includegraphics[width=.30\textwidth, clip=true, trim= 30px 40px 30px 50px]{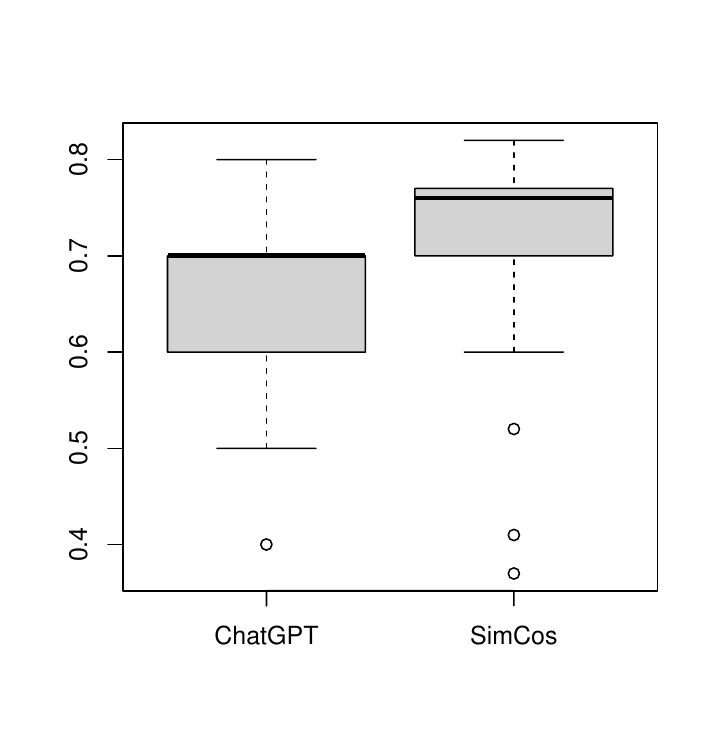}&
  \includegraphics[width=.30\textwidth, clip=true, trim= 30px 40px 30px 50px]{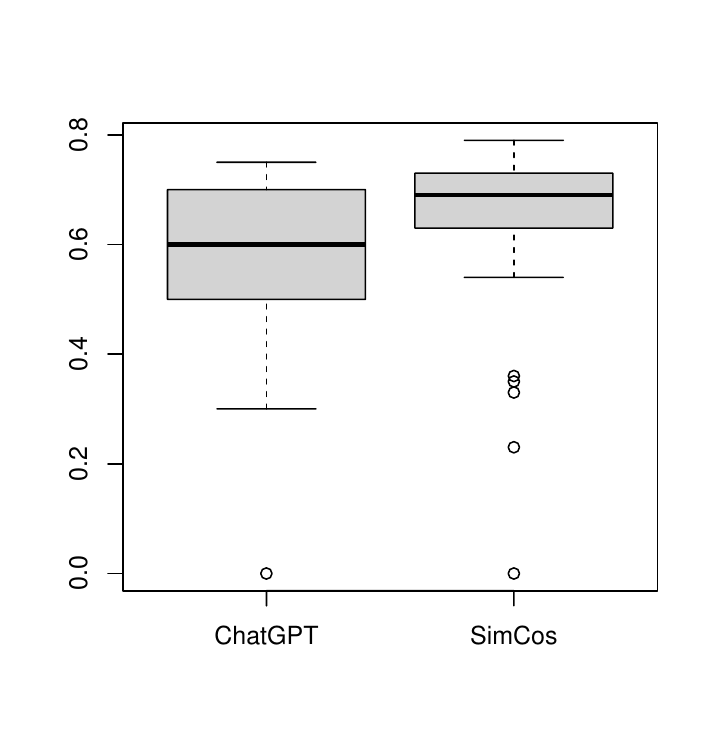}\\
  \textsc{Q4} & \textsc{Q5} & \textsc{Q6}\\
  \includegraphics[width=.30\textwidth, clip=true, trim= 30px 40px 30px 50px]{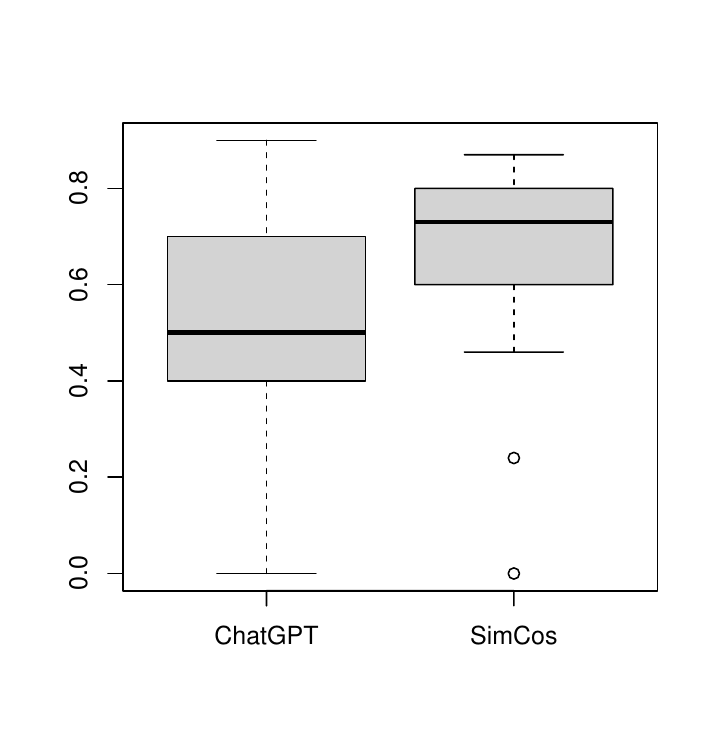}&
  \includegraphics[width=.30\textwidth, clip=true, trim= 30px 40px 30px 50px]{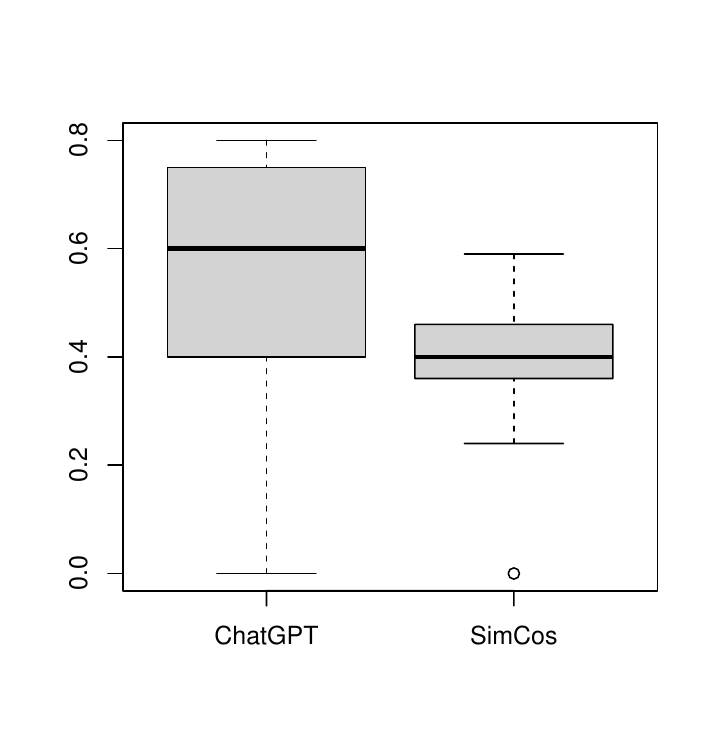}&
  \includegraphics[width=.30\textwidth, clip=true, trim= 30px 40px 30px 50px]{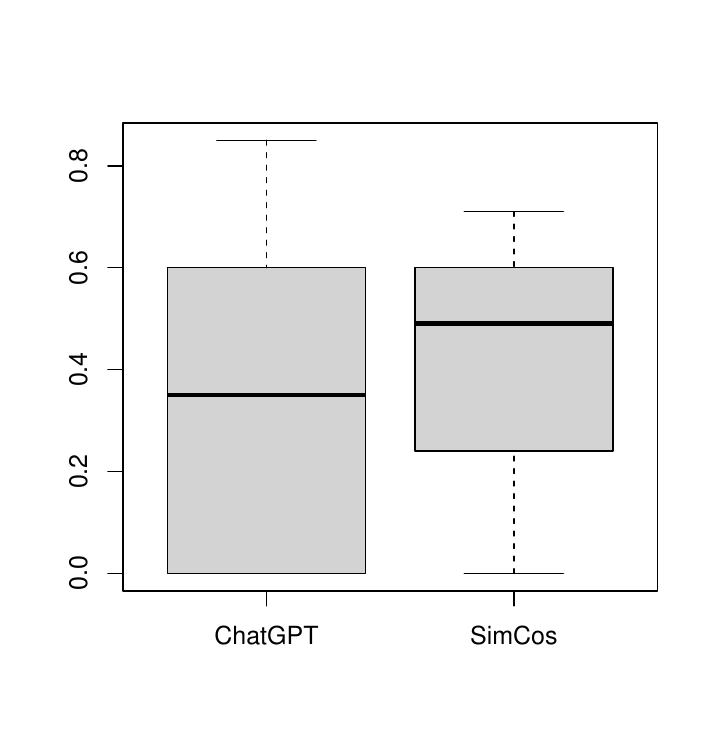}\\
\end{array}
$
\end{center}
\caption{Distribution of ChatGPT corrections and the cosine similarity metric.}
\label{fig:boxplot}
\end{figure*}

\vspace{0.2cm}
\noindent
\textbf{Reliability in ChatGPT's corrections.} As a way to complement the corrections provided by ChatGPT, we calculated the cosine similarity metric, which, in summary, evaluates the degree of proximity between two sentences in a vector space (further details in Section~\ref{sec:metrica}). Figure~\ref{fig:respostas} presents a comparison between ChatGPT's correction (normalized to range from 0 to 1) and the result of the cosine similarity metric calculation. In general, it can be observed that, for most questions, the lines follow a similar pattern: when one line tends to go up or down, the other line tends to follow suit.
The main difference in this trend can be observed in Q5 (Figure~\ref{fig:respostas}-(e)). In this particular question, on average, ChatGPT's scores had a mean of 5.25 points, while the mean of the similarity metric was 0.38. Figure~\ref{fig:boxplot} presents the distribution of corrections using ChatGPT and the cosine similarity metric. As observable, although the distribution of some responses has slightly different shapes, the values of mean and median remain close, with Q5 being the exception.

\vspace{0.2cm}
\noindent
\textbf{Divergences observed in ChatGPT's corrections.} To better understand the divergence observed in Q5, we investigated which responses had the greatest discrepancy in scores. We found a total of 10 responses that had more than 3 points of divergence. Upon analyzing these responses, we noticed that ChatGPT was able to detect details in the answers that potentially influenced the generated scores. For example, one participant answered Q5 as follows: ``\emph{response time, quantity of data transferred, and success rate (throughput) and error rate}''. However, ChatGPT noticed the confusion the participant made between success rate and throughput: ``\emph{The student's response addresses the main metrics, with some inaccuracies. He mentions response time, throughput, and error rate but confuses the quantity of data transferred with throughput. To clarify, throughput generally refers to the number of operations per unit of time and not the amount of data transferred. [...]}''. In other words, although the participant used the correct terms, the semantics of the terms were incorrect. This may have misled the cosine similarity metric, even though ChatGPT was able to detect and correct it.
Other cases of divergence occurred when the participant answered the question in a limited or incomplete manner. In these cases, again, the response contained certain terms used in the expert's response but lacked details or elaboration.

\section{Discussion: LLMs for Education}

The popularization of AI assistants could significantly impact traditional education. For example, students can use AI assistants to get answers to specific questions, clarify concepts or receive real-time feedback on their work. They are also auxiliary equity tools, considering the individual journeys of students. Neurodivergent people often cannot finish reading long texts, or they can at a high emotional cost and cognitive overload, something many neurotypical people do not imagine exists. ChatGPT and the like can help extract key points from readings, minimizing such effort. Teachers can benefit from the aid of these tools in creating teaching materials and personalizing instruction to meet individual student needs.

However, AI assistants should not replace the key role of teachers. Education goes beyond the transmission of information and involves social interactions, the development of socio-emotional skills and the ability to apply knowledge in real contexts. These aspects are fundamental and cannot be replaced by technology.

Finally, it is necessary to consider the ethical and privacy issues related to the use of AI assistants in education. It is important to ensure that student data is protected and that decisions related to education are not based solely on algorithms, but rather on a balanced approach that takes into account the expertise of teachers and the well-being of students. 

\section{Limitations}

As any empirical study, this work also has several limitations and threats to validity. First, to conduct the study, we used a population of 40 developers. Although they are professionals in the field, we can hardly generalize their responses to other groups of developers from different countries with different experiences.

Another threat is related to the feedback provided by ChatGPT. In some responses, we noticed that ChatGPT seems to have been too strict. For example, the question "What are the main metrics used to evaluate the performance of an application during a load test?" does not address potential relationships among these metrics. However, we frequently observed responses where ChatGPT provided feedback indicating this absence. Due to its proprietary implementation, the authors are not aware of the reasons why ChatGPT did not stick to answering what was asked.

Still, there is limitation related to the metric we used to compare ChatGPT's responses. Although there are other well-known metrics like BLEU~\cite{bleu} and METEOR~\cite{meteor}, we decided not to report them in this work due to inconsistent results we obtained. For example, a significant number of answers were evaluated as zero, which indicates that the student's answer does not match the reference answer. Moreover, BLEU is a metric designed to evaluate translations of texts, and we understand that the scenario of the article is not necessarily the most suitable for its use.
Other studies have observed that using these metrics is not appropriate for educational purposes as they do not correlate with human evaluation~\cite{scialom2019ask}.

\section{Related Work}
Although very recent, there is a growing number of studies using ChatGPT for educational purposes. For instance, the study conducted by Moore \textit{et al.}~\cite{moore2022assessing} explored the use of ChatGPT-3 to assess whether student-generated questions can be useful in the learning process. The results suggest that the model can be a powerful tool to assist teachers in pedagogical assessments, providing an innovative and effective approach to evaluate students' knowledge.

Zhu, Liu, and Lee~\cite{zhu2020effect} investigated the impact of automated assessment technologies on the review of scientific arguments presented by students. The results revealed that automated reviews were positively correlated with an increase in students' grades. Furthermore, when the automatic reviews were contextualized with the students' responses, they proved to be even more effective in aiding learning by providing personalized and specific feedback.

Additionally, Bernius, Krusche, and Bruegge~\cite{bernius2022machine} evaluated the use of LLMs to generate feedback for open-ended questions in courses with a large number of students. The results demonstrated a decrease of up to $85\%$ in the effort required by teachers for evaluating these questions. Furthermore, $92\%$ of the evaluations generated by LLMs were considered of high quality by instructors. Moreover, the majority of students rated the quality of this feedback as equivalent to that provided by instructors.

These studies demonstrate how the use of ChatGPT and automated assessment technologies can have a positive impact on the learning process of students. These tools offer the opportunity to provide immediate, personalized, and detailed feedback, allowing students to improve their performance and understanding of concepts. However, none of these works addressed the use of tools like ChatGPT for the correction and feedback of open-ended questions in the field of software engineering, which is the objective of this study.

%Além disso, elas podem aliviar a carga de trabalho dos professores e instrutores ao automatizar parte do processo de avaliação, liberando tempo para outras atividades educacionais.

\section{Conclusions}

This study investigated the use of ChatGPT as a complementary strategy for correcting and providing feedback on open-ended questions. For this purpose, we recruited two experts and 40 developers to answer a set of six questions on two different topics: (1) caching and (2) stress and performance testing. The responses from these individuals were corrected by ChatGPT. Based on this data, several observations were made: 

\begin{itemize}
    \item In general, experts agreed with the corrections provided by ChatGPT. Among the six feedbacks given by ChatGPT, there was only one instance of disagreement from the expert;

    \item The cosine similarity metric is not always suitable to be used as a proxy for ChatGPT scores, as it loses contextual information that ChatGPT is able to identify.

\end{itemize}

\subsection{Future Work} 

For future work, we are interested in using grading rubrics annotated by experts with specific weights for certain items that deserve emphasis in the responses. This way, we can direct corrections to the most relevant response items. We also hope to expand our analysis to cover other LLMs, as well as different versions of ChatGPT. 
Moreover, we plan to explore other teaching materials, such as books, as our database for open questions (and their answers).  This approach may further enrich the analysis and offer a more comprehensive view on the effectiveness and applicability of ChatGPT in education.
Finally, we still want to understand to what extend does ChatGPT recognizes its own answers and, thus, give better grades to them. 

\subsection*{Artifacts Availability}
The data analyzed in this study is available online at: \url{https://tinyurl.com/chatgpt-for-edu}.

\subsection*{Acknowledgements}
We thank all the Zuppers who answered the questionnaire and the reviewers who provided relevant suggestions for improvements. This work is partially supported by FAPESPA (\#053/2021) and CNPq (\#308623/2022-3).

\subsection*{AI Tooling}

Ultimately, although this work used ChatGPT as a subject under study, we did not use ChatGPT to generate text content for this paper. However, this paper was initially submitted in Portuguese, but later translated into English with the support of ChatGPT (using the prompt ``The text below is written using the LateX markup language. Translate it to English, so it keeps the original LaTeX markup. Do not perform any other textual adjustments.'').

\bibliographystyle{ACM-Reference-Format}
\bibliography{references}
\end{document}

%% file: packages.tex
\usepackage{epsfig}
\usepackage{subfigure}
\usepackage{calc}
\usepackage{amstext}
\usepackage{amsmath}
\usepackage{amsthm}
\usepackage{multicol}
\usepackage{pslatex}
\usepackage{multirow}
\usepackage{booktabs}
\usepackage{tikz}
\usepackage{graphicx}
\usepackage{xcolor}
\usepackage{url}
\usepackage[T1]{fontenc}
\usepackage[utf8]{inputenc}
\usepackage{array}
\usepackage{xspace}